\input harvmac
\noblackbox

\input epsf
   
 
\def\journal#1&#2(#3){\unskip, \sl #1\ \bf #2 \rm(19#3) }
\def\andjournal#1&#2(#3){\sl #1~\bf #2 \rm (19#3) }

\def\frac#1#2{{#1\over#2}}

\def\inbar{\,\vrule height1.5ex width.4pt depth0pt}
\def\IC{\relax\hbox{$\inbar\kern-.3em{\rm C}$}}
\def\IR{\relax{\rm I\kern-.18em R}}
\def\IP{\relax{\rm I\kern-.18em P}}
\def\IZ{\relax{\rm I\kern-.18em Z}}
\def\IE{\relax{\rm I\kern-.18em E}}
\def\IT{\relax{\rm I\kern-.44em T}}

%
%

%
\catcode`\@=11
\def\slash#1{\mathord{\mathpalette\c@ncel{#1}}}
\overfullrule=0pt

\def\II{{\cal I}}

\def\NN{{\cal N}}

\def\ZZ{{\cal Z}}

\def\underrel#1\over#2{\mathrel{\mathop{\kern\z@#1}\limits_{#2}}}

\catcode`\@=12


%


\def\i{{\bf i}}


\def\unlockat{\catcode`\@=11}
\def\lockat{\catcode`\@=12}

\unlockat


\def\newsec#1{\global\advance\secno by1\message{(\the\secno. #1)}
\global\subsecno=0\global\subsubsecno=0\eqnres@t\noindent
{\bf\the\secno. #1}
\writetoca{{\secsym} {#1}}\par\nobreak\medskip\nobreak}
\global\newcount\subsecno \global\subsecno=0
\def\subsec#1{\global\advance\subsecno
by1\message{(\secsym\the\subsecno. #1)}
\ifnum\lastpenalty>9000\else\bigbreak\fi\global\subsubsecno=0
\noindent{\it\secsym\the\subsecno. #1}
\writetoca{\string\quad {\secsym\the\subsecno.} {#1}}
\par\nobreak\medskip\nobreak}
\global\newcount\subsubsecno \global\subsubsecno=0
\def\subsubsec#1{\global\advance\subsubsecno by1
\message{(\secsym\the\subsecno.\the\subsubsecno. #1)}
\ifnum\lastpenalty>9000\else\bigbreak\fi
\noindent\quad{\secsym\the\subsecno.\the\subsubsecno.}{#1}
\writetoca{\string\qquad{\secsym\the\subsecno.\the\subsubsecno.}{#1}}
\par\nobreak\medskip\nobreak}

\def\subsubseclab#1{\DefWarn#1\xdef
#1{\noexpand\hyperref{}{subsubsection}%
{\secsym\the\subsecno.\the\subsubsecno}%
{\secsym\the\subsecno.\the\subsubsecno}}%
\writedef{#1\leftbracket#1}\wrlabeL{#1=#1}}
\lockat


\newcount\figno
\figno=1
\def\fig#1#2#3{
\par\begingroup\parindent=0pt\leftskip=1cm\rightskip=1cm\parindent=0pt
\baselineskip=11pt
\global\advance\figno by 1
\midinsert
\epsfxsize=#3
\centerline{\epsfbox{#2}}
{\bf Fig.\ \the\figno: } #1\par
\endinsert\endgroup\par
}
\def\figlabel#1{\xdef#1{\the\figno}}
\def\encadremath#1{\vbox{\hrule\hbox{\vrule\kern8pt\vbox{\kern8pt
\hbox{$\displaystyle #1$}\kern8pt}
\kern8pt\vrule}\hrule}}
%
%


\font\cmss=cmss10
\font\cmsss=cmss10 at 7pt
\def\rlx{\relax\leavevmode}
\def\inbar{\vrule height1.5ex width.4pt depth0pt}
\def\IC{\relax\,\hbox{$\inbar\kern-.3em{\rm C}$}}
\def\IN{\relax{\rm I\kern-.18em N}}
\def\IP{\relax{\rm I\kern-.18em P}}
\def\ZZ{\rlx\leavevmode\ifmmode\mathchoice{\hbox{\cmss Z\kern-.4em Z}}
 {\hbox{\cmss Z\kern-.4em Z}}{\lower.9pt\hbox{\cmsss Z\kern-.36em Z}}
 {\lower1.2pt\hbox{\cmsss Z\kern-.36em Z}}\else{\cmss Z\kern-.4em
 Z}\fi}
\def\IZ{\relax\ifmmode\mathchoice
{\hbox{\cmss Z\kern-.4em Z}}{\hbox{\cmss Z\kern-.4em Z}}
{\lower.9pt\hbox{\cmsss Z\kern-.4em Z}}
{\lower1.2pt\hbox{\cmsss Z\kern-.4em Z}}\else{\cmss Z\kern-.4em
Z}\fi}
\def\IZ{\relax\ifmmode\mathchoice
{\hbox{\cmss Z\kern-.4em Z}}{\hbox{\cmss Z\kern-.4em Z}}
{\lower.9pt\hbox{\cmsss Z\kern-.4em Z}}
{\lower1.2pt\hbox{\cmsss Z\kern-.4em Z}}\else{\cmss Z\kern-.4em
Z}\fi}

\def\narrowplus{\kern -.04truein + \kern -.03truein}
\def\narrowminus{- \kern -.04truein}
\def\narrowminussub{\kern -.02truein - \kern -.01truein}

\def\IZ{\relax\ifmmode\mathchoice
{\hbox{\cmss Z\kern-.4em Z}}{\hbox{\cmss Z\kern-.4em Z}}
{\lower.9pt\hbox{\cmsss Z\kern-.4em Z}}
{\lower1.2pt\hbox{\cmsss Z\kern-.4em Z}}\else{\cmss Z\kern-.4em
Z}\fi}
\def\IB{\relax{\rm I\kern-.18em B}}
\def\IC{{\relax\hbox{$\inbar\kern-.3em{\rm C}$}}}
\def\ID{\relax{\rm I\kern-.18em D}}
\def\IE{\relax{\rm I\kern-.18em E}}
\def\IF{\relax{\rm I\kern-.18em F}}
\def\IG{\relax\hbox{$\inbar\kern-.3em{\rm G}$}}
\def\IGa{\relax\hbox{${\rm I}\kern-.18em\Gamma$}}
\def\IH{\relax{\rm I\kern-.18em H}}
\def\II{\relax{\rm I\kern-.18em I}}
\def\IK{\relax{\rm I\kern-.18em K}}
\def\IP{\relax{\rm I\kern-.18em P}}

\font\cmss=cmss10 \font\cmsss=cmss10 at 7pt
\def\IR{\relax{\rm I\kern-.18em R}}


%

%
%
\def\eqnn#1{\xdef #1{(\secsym\the\meqno)}\writedef{#1\leftbracket#1}%
\global\advance\meqno by1\wrlabeL#1}
\def\eqna#1{\xdef #1##1{\hbox{$(\secsym\the\meqno##1)$}}
\writedef{#1\numbersign1\leftbracket#1{\numbersign1}}%
\global\advance\meqno by1\wrlabeL{#1$\{\}$}}
\def\eqn#1#2{\xdef #1{(\secsym\the\meqno)}\writedef{#1\leftbracket#1}%
\global\advance\meqno by1$$#2\eqno#1\eqlabeL#1$$}


\def\boxit#1{\vbox{\hrule\hbox{\vrule\kern8pt
\vbox{\hbox{\kern8pt}\hbox{\vbox{#1}}\hbox{\kern8pt}}
\kern8pt\vrule}\hrule}}
\def\mathboxit#1{\vbox{\hrule\hbox{\vrule\kern5pt\vbox{\kern5pt
\hbox{$\displaystyle #1$}\kern5pt}\kern5pt\vrule}\hrule}}


\lref\GiveonZN{
  A.~Giveon and D.~Kutasov,
  ``Seiberg Duality in Chern-Simons Theory,''
Nucl.\ Phys.\ B {\bf 812}, 1 (2009).
[arXiv:0808.0360 [hep-th]].
}

\lref\bult{
 F.~van de Bult, 
 ``Hyperbolic Hypergeometric Functions,''
{http://dare.uva.nl/document/97725}.
} 

\lref\BeniniMF{
  F.~Benini, C.~Closset and S.~Cremonesi,
  ``Comments on 3d Seiberg-like dualities,''
JHEP {\bf 1110}, 075 (2011).
[arXiv:1108.5373 [hep-th]].
}

\lref\WillettGP{
  B.~Willett and I.~Yaakov,
  ``N=2 Dualities and Z Extremization in Three Dimensions,''
[arXiv:1104.0487 [hep-th]].
}

\lref\Rainslimits{
   E.~M.~Rains,
   ``Limits of elliptic hypergeometric integrals,''
[arXiv:math.CA/0607093].
}   

\lref\Rainstrans{
   E.~M.~Rains,
   ``Transformations of elliptic hypergeometric integrals,''
Ann.~Math.~Vol.~171, 1 (2010)
}

\lref\DolanQI{
  F.~A.~Dolan and H.~Osborn,
  ``Applications of the Superconformal Index for Protected Operators and q-Hypergeometric Identities to N=1 Dual Theories,''
Nucl.\ Phys.\ B {\bf 818}, 137 (2009).
[arXiv:0801.4947 [hep-th]].
}

\lref\SpiridonovZA{
  V.~P.~Spiridonov and G.~S.~Vartanov,
  ``Elliptic Hypergeometry of Supersymmetric Dualities,''
Commun.\ Math.\ Phys.\  {\bf 304}, 797 (2011).
[arXiv:0910.5944 [hep-th]].
}

\lref\MoritaCS{
  T.~Morita and V.~Niarchos,
  ``F-theorem, duality and SUSY breaking in one-adjoint Chern-Simons-Matter theories,''
Nucl.\ Phys.\ B {\bf 858}, 84 (2012).
[arXiv:1108.4963 [hep-th]].
}

\lref\SeibergNZ{
  N.~Seiberg and E.~Witten,
  ``Gauge dynamics and compactification to three-dimensions,''
In *Saclay 1996, The mathematical beauty of physics* 333-366.
[hep-th/9607163].
}

\lref\DolanRP{
  F.~A.~H.~Dolan, V.~P.~Spiridonov and G.~S.~Vartanov,
  ``From 4d superconformal indices to 3d partition functions,''
Phys.\ Lett.\ B {\bf 704}, 234 (2011).
[arXiv:1104.1787 [hep-th]].
}

\lref\NiarchosJB{
  V.~Niarchos,
  ``Seiberg Duality in Chern-Simons Theories with Fundamental and Adjoint Matter,''
JHEP {\bf 0811}, 001 (2008).
[arXiv:0808.2771 [hep-th]].
}

\lref\NiarchosAA{
  V.~Niarchos,
  ``R-charges, Chiral Rings and RG Flows in Supersymmetric Chern-Simons-Matter Theories,''
JHEP {\bf 0905}, 054 (2009).
[arXiv:0903.0435 [hep-th]].
}

\lref\RomelsbergerEC{
  C.~Romelsberger,
  ``Calculating the Superconformal Index and Seiberg Duality,''
[arXiv:0707.3702 [hep-th]].
}

\lref\GaddeIA{
  A.~Gadde and W.~Yan,
  ``Reducing the 4d Index to the $S^3$ Partition Function,''
[arXiv:1104.2592 [hep-th]].
}

\lref\ImamuraUW{
  Y.~Imamura,
  ``Relation between the 4d superconformal index and the $S^3$ partition function,''
JHEP {\bf 1109}, 133 (2011).
[arXiv:1104.4482 [hep-th]].
}

\lref\ImamuraWG{
  Y.~Imamura and D.~Yokoyama,
  ``N=2 supersymmetric theories on squashed three-sphere,''
Phys.\ Rev.\ D {\bf 85}, 025015 (2012).
[arXiv:1109.4734 [hep-th]].
}

\lref\HamaAV{
  N.~Hama, K.~Hosomichi and S.~Lee,
  ``Notes on SUSY Gauge Theories on Three-Sphere,''
JHEP {\bf 1103}, 127 (2011).
[arXiv:1012.3512 [hep-th]].
}

\lref\HamaEA{
  N.~Hama, K.~Hosomichi and S.~Lee,
  ``SUSY Gauge Theories on Squashed Three-Spheres,''
JHEP {\bf 1105}, 014 (2011).
[arXiv:1102.4716 [hep-th]].
}

\lref\JafferisUN{
  D.~L.~Jafferis,
  ``The Exact Superconformal R-Symmetry Extremizes Z,''
[arXiv:1012.3210 [hep-th]].
}

\lref\KapustinKZ{
  A.~Kapustin, B.~Willett and I.~Yaakov,
  ``Exact Results for Wilson Loops in Superconformal Chern-Simons Theories with Matter,''
JHEP {\bf 1003}, 089 (2010).
[arXiv:0909.4559 [hep-th]].
}

\lref\IntriligatorNE{
  K.~A.~Intriligator and P.~Pouliot,
  ``Exact superpotentials, quantum vacua and duality in supersymmetric SP(N(c)) gauge theories,''
Phys.\ Lett.\ B {\bf 353}, 471 (1995).
[hep-th/9505006].
}

\lref\SeibergPQ{
  N.~Seiberg,
  ``Electric - magnetic duality in supersymmetric nonAbelian gauge theories,''
Nucl.\ Phys.\ B {\bf 435}, 129 (1995).
[hep-th/9411149].
}

\lref\AharonyGP{
  O.~Aharony,
  ``IR duality in d = 3 N=2 supersymmetric USp(2N(c)) and U(N(c)) gauge theories,''
Phys.\ Lett.\ B {\bf 404}, 71 (1997).
[hep-th/9703215].
}

\lref\SpiridonovQV{
  V.~P.~Spiridonov and G.~S.~Vartanov,
  ``Superconformal indices of ${\NN}=4$ SYM field theories,''
Lett.\ Math.\ Phys.\  {\bf 100}, 97 (2012).
[arXiv:1005.4196 [hep-th]].
}

\lref\SpiridonovEM{
  V.~P.~Spiridonov,
  ``Elliptic beta integrals and solvable models of statistical mechanics,''
[arXiv:1011.3798 [hep-th]].
}

\lref\KapustinXQ{
  A.~Kapustin, B.~Willett and I.~Yaakov,
  ``Nonperturbative Tests of Three-Dimensional Dualities,''
JHEP {\bf 1010}, 013 (2010).
[arXiv:1003.5694 [hep-th]].
}

\lref\KapustinMH{
  A.~Kapustin, B.~Willett and I.~Yaakov,
  ``Tests of Seiberg-like Duality in Three Dimensions,''
[arXiv:1012.4021 [hep-th]].
}

\lref\KutasovVE{
  D.~Kutasov,
  ``A Comment on duality in N=1 supersymmetric nonAbelian gauge theories,''
Phys.\ Lett.\ B {\bf 351}, 230 (1995).
[hep-th/9503086].
}

\lref\BrodieVX{
  J.~H.~Brodie,
  ``Duality in supersymmetric SU(N(c)) gauge theory with two adjoint chiral superfields,''
Nucl.\ Phys.\ B {\bf 478}, 123 (1996).
[hep-th/9605232].
}

\lref\Spiridonov{
 V.~P.~Spiridonov,
 ``Essays on the theory of elliptic hypergeometric functions,''
 Ru. Math. Surveys, Volume 63, Issue 3, 405-472 (2008).
 [arXiv:0805.3135 [math.CA]].
 }

\lref\Diejen{
 J.~F.~ van Diejen and V.~P.~Spiridonov,
  ``Unit circle elliptic beta integrals,''
  Ramanujan Journal 10, no 2, 187-204 (2005)
  [arXiv:math/0309279].
}  

\lref\Spiridonova{
 V.~P.~Spiridonov, 
 ``On the elliptic beta function,'' 
 Uspekhi Mat. Nauk {\bf 56} (1) 181-182 (2001),  
 (Russian Math. Surveys {\bf 56} (1) 185-186 (2001)).
}  

\lref\Spiridonovb{
  V.~P.~Spiridonov, 
  ``Theta hypergeometric integrals,'' 
  Algebra i Analiz {\bf 15} (6) 161-215 (2003), 
  (St.Petersburg Math. J. {\bf 15} (6) (2004), 929--967) 
  [arXiv:math/0303205].
}



\rightline{CCTP-2012-07}
\vskip 16pt
\Title{
}
{\vbox{\centerline{Seiberg dualities and the 3d/4d connection}
}}
\medskip
\centerline{Vasilis Niarchos}
\bigskip
\centerline{{\it Crete Center for Theoretical Physics}}
\centerline{\it Department of Physics, University of Crete, 71303, Greece}
\bigskip
\centerline{niarchos@physics.uoc.gr}
\bigskip\bigskip\bigskip
\centerline{\bf Abstract}
\bigskip

\noindent
We discuss the degeneration limits of $d=4$ superconformal indices that relate
Seiberg duality for the $d=4$ $\NN=1$ SQCD theory to Aharony and Giveon-Kutasov dualities
for $d=3$ $\NN=2$ SQCD theories. On a mathematical level we argue that this 3d/4d
connection entails a new set of non-standard degeneration identities between 
hyperbolic hypergeometric integrals. On a physical level we propose that such
degeneration formulae provide a new route to the still illusive Seiberg
dualities for $d=3$ $\NN=2$ SQCD theories with $SU(N)$ gauge group.

\vfill
\Date{}



\newsec{Degeneration schemes of partition functions and the 3d/4d connection}
\seclab\intro

Quantum field theories (QFTs) related by reduction on a spatial $S^1$ frequently exhibit similar
properties, $e.g.$ similarities in duality and spontaneous supersymmetry breaking patterns. 
One can try to trace the quantum dynamics of the compactified theory as a function of the
compactification radius \SeibergNZ, but this is typically hard.

The superconformal indices (SCIs) of $d=4$ QFTs provide an interesting new perspective on such 
relations. Under an $S^1$ reduction the SCI, which is a partition function on $S^3\times S^1$,
reduces to the $S^3$ partition function of a three dimensional QFT 
\refs{\DolanRP\GaddeIA\ImamuraUW-\ImamuraWG}. For generic gauge theories
with a known UV Lagrangian description there is a standard prescription for the computation of 
$d=4$ SCIs and $S^3$ partition functions. The $d=4$ SCIs are expressed in terms of elliptic
hypergeometric integrals \DolanQI\ and the $S^3$ partition functions in terms of hyperbolic 
hypergeometric integrals \refs{\KapustinKZ\JafferisUN\HamaAV-\HamaEA}. Original work 
on the mathematics of the elliptic hypergeometric integrals was performed in 
\refs{\Spiridonova,\Spiridonovb} (see \Spiridonov\ for a review). A lengthy treatise on 
hyperbolic hypergeometric integrals, whose notation we will follow closely, is \bult. The 
first paper to describe the reduction from elliptic to hyperbolic hypergeometric integrals was \Diejen.

In this framework, a field theory duality in four dimensions translates to a corresponding 
duality transformation property of elliptic hypergeometric integrals. The subsequent reduction of
this transformation to hyperbolic hypergeometric integrals implies a corresponding field theory 
duality in three dimensions. It is believed that every duality in four dimensions \SpiridonovZA\ 
descends in this manner to a duality in three dimensions \DolanRP.

In practice, the descent between a four dimensional and a three dimensional duality identity 
is not just a single $S^1$ reduction of the four dimensional SCI but a sequence of 
reductions whose purpose is to remove constraining conditions on external parameters, $e.g.$ real 
masses, and/or add extra parameters like Fayet-Iliopoulos (FI) terms and Chern-Simons (CS) 
interactions. The latter steps are crucial at the end of the process when we read off the specifics of the 
three dimensional duality from the corresponding form of the duality transformation properties of 
hyperbolic hypergeometric integrals. Examples of such reductions in a field theory context have been 
provided in \refs{\SpiridonovQV,\DolanRP}. 

The mathematical implementation of these steps relies on specific degeneration schemes between
elliptic and/or hyperbolic hypergeometric integrals. Such schemes have been studied in the 
mathematics literature in \refs{\Rainslimits,\bult} and have been implemented in 
\refs{\SpiridonovQV,\DolanRP,\KapustinXQ\KapustinMH\WillettGP-\BeniniMF} to demonstrate 
certain $d=3$ dualities on the level of $S^3$ partition functions. In this note we will argue that the 
generic reduction between a $d=4$ and a $d=3$ duality involves more general degeneration 
schemes with qualitatively new features whose study is both physically and mathematically 
interesting.

For concreteness, in this paper we will focus on the example of $d=4$ Seiberg duality \SeibergPQ\ 
and its reduction to $d=3$ Aharony \AharonyGP\ and Giveon-Kutasov dualities \GiveonZN. 
The known route to the integral identities implied by the matching of $S^3$ partition functions in 
Aharony/Giveon-Kutasov dualities proceeds along the lines of the following degeneration scheme.
The starting point is Seiberg duality for the $d=4$ $\NN=1$ SQCD theory with gauge group 
$Sp(2N)$ (also known as the Intriligator-Pouliot duality \IntriligatorNE), and the corresponding 
transformation properties of SCIs, which were proven in \Rainstrans, are of the BC type. 
The $S^1$ degeneration of these identities 
becomes a duality transformation property of the so-called $I_{BC}$ top level integral.
A subsequent degeneration scheme \bult\ that reduces the $I_{BC}$ top level integral to the $S^3$
partition functions of $\NN=2$ SQCD and Chern-Simons SQCD theories with gauge group $U(N)$
allows the derivation of the transformation properties implied by Aharony/Giveon-Kutasov dualities. 

All the reductions involved in this particular degeneration scheme share the following 
(technically convenient) features: $(i)$ they keep the number of integration variables invariant, 
and $(ii)$ they can be derived by exchanging the integral with the degeneration limits.
In what follows we will call such reductions `standard'. We will argue that there are also more 
involved reductions that do not obey $(i)$ and $(ii)$. We will call the latter `non-standard reductions'.

We notice that the starting point of the above scheme is not Seiberg duality for the $d=4$ $\NN=1$
SQCD theory with $SU(N)$ gauge group. Since there is a direct $S^1$ reduction between 
the $d=4$ $\NN=1$ and $d=3$ $\NN=2$ SQCD theories with unitary gauge group it is physically 
more interesting to find a degeneration scheme between the partition functions of these theories.
This entails a more direct connection between the duality transformation properties of 
the $d=4$ $\NN=1$ $SU(N)$ SQCD theory \refs{\Rainstrans,\DolanQI}, 
and the transformation properties of hyperbolic hypergeometric integrals required by 
Aharony/Giveon-Kutasov dualities \bult.
Our main goal will be to discuss explicitly how this connection is implemented and what
mathematical properties it requires. We will find that 
by gauging the baryon symmetry of the $d=4$ SQCD theory 
we recover the Aharony/Giveon-Kutasov dualities for $U(N)$ $\NN=2$ SQCD theories.
Without gauging the baryon symmetry of the four dimensional theory we obtain a mathematically
concrete route towards a long-suspected 3d Seiberg duality for $SU(N)$ $\NN=2$ SQCD theories. 
The latter cannot be derived using the degeneration scheme that starts with the $d=4$ 
$Sp(2N)$ Intriligator-Pouliot duality. 

Analogous reduction schemes can be implemented for more general $d=4$ Seiberg dualities 
\SpiridonovZA. The descent between Kutasov \KutasovVE\ and Brodie \BrodieVX\ dualities with 
adjoint and fundamental matter to their $d=3$ descendants \refs{\NiarchosJB,\NiarchosAA} is an
example. We will not discuss explicitly this possibility in this paper.

\newsec{From $d=4$ Seiberg duality to $d=3$ Aharony/Giveon-Kutasov duality}

\subsec{The superconformal index of $d=4$ $\NN=1$ SQCD}

The SCI of the $d=4$ $\NN=1$ SQCD theory with $N_f$ pairs of quark supermultiplets
in the (anti)fundamental representation of the gauge group $SU(N_c)$ is 
\refs{\RomelsbergerEC,\DolanQI} 
(for the precise conventions used here see also eqs.\ (4.6), (4.7) of the review \SpiridonovZA):
\eqn\sciaa{\eqalign{
&I_E^{(SU)}(N_c,N_f;s;t)=\frac{(p;p)_\infty^{N_c-1} (q;q)_\infty^{N_c-1}}{N_c!}
\cr
&\int_{\IT^{N_c-1}} 
\prod_{j=1}^{N_c-1} \frac{dz_j}{2\pi \i z_j} \,
\frac{\displaystyle \prod_{a=1}^{N_f} \prod_{j=1}^{N_c} \Gamma_e(s_a z_j,t_a^{-1}z_j^{-1};p,q)}
{\displaystyle \prod_{1\leq i<j\leq N_c}\Gamma_e(z_i z_j^{-1},z_i^{-1}z_j;p,q)} 
\Bigg |_{\prod_{j=1}^{N_c} z_j=1}
}}
for the electric description, and
\eqn\sciab{\eqalign{
&I_M^{(SU)}(\tilde N_c,N_f;s;t)=\frac{(p;p)_\infty^{\tilde N_c-1}(q;q)^{\tilde N_c-1}_\infty}{\tilde N_c!} 
\prod_{a,b=1}^{N_f} \Gamma_e(s_a t_b^{-1};p,q)
\cr
&\int_{\IT^{\tilde N_c-1}} 
\prod_{j=1}^{\tilde N_c-1} \frac{dz_j}{2\pi \i z_j} \,
\frac{\displaystyle \prod_{a=1}^{N_f} \prod_{j=1}^{\tilde N_c} 
\Gamma_e(S^{\frac{1}{\tilde N_c}}s_a^{-1} z_j,T^{-\frac{1}{\tilde N_c}} t_az_j^{-1};p,q)}
{\displaystyle \prod_{1\leq i<j\leq \tilde N_c}\Gamma_e(z_i z_j^{-1},z_i^{-1}z_j;p,q)} 
\Bigg |_{\prod_{j=1}^{\tilde N_c} z_j=1}
}}
for the magnetic description. 

We make a short parenthesis to explain the notation. The rank of the dual gauge group will be 
denoted as 
$$
\tilde N_c=N_f-N_c
~.$$
$(z,p)_\infty$ is the $q$-Pochhammer symbol (thus $(p,p)_\infty$ is equivalent to the 
Euler function $\phi(p)$) and $\Gamma_e(z;p,q)$ the elliptic $\Gamma_e$-function (we refer the 
reader to \refs{\SpiridonovZA,\bult} for precise definitions and references to the original literature). 
We use the common convention $\Gamma_e(z_1,z_2;p,q)=\Gamma_e(z_1;p,q)\Gamma_e(z_2;p,q)$. 
In the expressions \sciaa, \sciab\ the external vector 
parameters $s=(s_1,\ldots,s_{N_f})$ and $t=(t_1,\ldots,t_{N_f})$ denote (renormalized) 
fugacities of the global flavor group, whereas $p,q$ are fugacities related to the U(1) 
$R$-symmetry of the theory (further details are available in the review \SpiridonovZA). 
The fugacities obey the balancing conditions
\eqn\sciac{
S := \prod_{a=1}^{N_f} s_a=(pq)^{N_f r_Q}~, ~~
T := \prod_{a=1}^{N_f} t_a=(pq)^{-N_f r_{\tilde Q}}
~}
where 
\eqn\sciaca{
r_Q=\frac{\tilde N_c}{2N_f}+x~, ~~ r_{\tilde Q}=\frac{{\tilde N_c}}{2N_f}-x
}
are the $R$-charges of the fundamental and antifundamental multiplets $Q$, $\tilde Q$. $x$ captures
the effects of a baryon $U(1)_B$ fugacity.

Seiberg duality implies the following mathematical identity
\eqn\sciae{
I_E^{(SU)}(N_c,N_f;s;t)=I_M^{(SU)}(\tilde N_c, N_f;s;t)
~.}
It was shown by Dolan and Osborn \DolanQI\ that this identity coincides with the 
$A_n\leftrightarrow A_m$ root systems symmetry transformation established by Rains in 
\Rainstrans.

By gauging the baryon symmetry $U(1)_B$ it is not difficult to derive the $U(N)$ version of
the identity \sciae\
\eqn\sciaf{
I_E^{(U)}(N_c,N_f;s;t)=I_M^{(U)}(\tilde N_c, N_f;s,t)
}
where
\eqn\sciag{\eqalign{
I_E^{(U)}(N_c,N_f;s;t)&=\frac{(p;p)_\infty^{N_c-1}(q;q)^{N_c-1}_\infty}{N_c!} 
\int_{\IT^{N_c}} \prod_{j=1}^{N_c} \frac{dz_j}{2\pi \i z_j}
\frac{\displaystyle \prod_{a=1}^{N_f} \prod_{j=1}^{N_c} \Gamma_e(s_a z_j,t_a^{-1}z_j^{-1};p,q)}
{\displaystyle \prod_{1\leq i<j\leq N_c}\Gamma_e(z_i z_j^{-1},z_i^{-1}z_j;p,q)} 
\cr
&=\int_{S^1}\frac{dx}{2\pi \i x} I_E^{(SU)}(N_c, N_f;x^{-1} s; x^{-1} t)
~,}}
and
\eqn\sciai{\eqalign{
&I_M^U(\tilde N_c, N_f;s;t)=\frac{(p;p)_\infty^{\tilde N_c-1}(q;q)^{\tilde N_c-1}_\infty}{\tilde N_c!} 
\prod_{a,b=1}^{N_f} \Gamma_e(s_a t_b^{-1};p,q)
\cr
&\int_{\IT^{\tilde N_c}} \prod_{j=1}^{\tilde N_c} \frac{dz_j}{2\pi \i z_j}
\frac{\displaystyle \prod_{a=1}^{N_f} \prod_{j=1}^{\tilde N_c} 
\Gamma_e(S^{\frac{1}{\tilde N_c}}s_a^{-1} z_j,T^{-\frac{1}{\tilde N_c}} t_a z_j^{-1};p,q)}
{\displaystyle \prod_{1\leq i<j\leq \tilde N_c}\Gamma_e(z_i z_j^{-1},z_i^{-1}z_j;p,q)} 
=\int_{S^1}\frac{dx}{2\pi \i x} I_M^{(SU)}(\tilde N_c, N_f;x^{-1} s; x^{-1} t)
.}}

In what follows we consider a degeneration scheme based on the $U(N)$ identity \sciaf. We will 
return to the reduction of the $SU(N)$ identity \sciae\ in section 4.

\subsec{First degeneration: the $S^1$ reduction}

Following standard procedure we set
\eqn\redaa{
p=e^{2\pi \i v\omega_1}~, ~~ q=e^{2\pi \i v\omega_2}~, ~~
s_a=e^{2\pi \i v \mu_a}~,~~ t_a=e^{-2\pi \i v \nu_a}~, ~~ z_j=e^{2\pi \i v u_j}
~,}
where $i,j=1,\ldots,N_c$, $a=1,\ldots,N_f$, and take the degeneration limit $v\to 0$.
In this limit the elliptic $\Gamma_e$-functions reduce to hyperbolic $\Gamma_h$-functions
and by exchanging limit and integral we obtain the degeneration formulae
\eqn\redab{
\lim_{v\to 0} I_E^{(U)}(N_c,N_f;s,t)=\sqrt{-v^2 \omega_1\omega_2}\, 
e^{\frac{\pi \i \omega (N_c^2+1)}{6v\omega_1\omega_2}}\,
J_{N_c,(N_f,N_f),0}(\mu;\nu;0)
~,}
\eqn\redac{\eqalign{
\lim_{v\to 0} I_M^{(U)}(\tilde N_c,N_f;s,t)=&\sqrt{-v^2 \omega_1\omega_2}\,
e^{\frac{\pi \i \omega (N_c^2+1)}{6v\omega_1\omega_2}}
\prod_{a,b=1}^{N_f} \Gamma_h(\mu_a+\nu_b;\omega_1,\omega_2)\,
\cr
&J_{\tilde N_c,(N_f,N_f),0}(\omega-\nu;\omega-\mu;0)
~,}}
where $J_{n,(s_1,s_2),t}$ is the function 
\eqn\redad{\eqalign{
&J_{n,(s_1,s_2),t}(\mu;\nu;2\lambda)=
\frac{1}{n!}\int \prod_{j=1}^n\left( \frac{du_j}{\sqrt{-\omega_1\omega_2}}
e^{\frac{2\pi \i \lambda u_j}{\omega_1\omega_2}}
e^{\frac{\pi \i t u_j^2}{2\omega_1\omega_2}}\right)
\cr
&\frac{\displaystyle \prod_{j=1}^n \prod_{a=1}^{s_1}
\Gamma_h(\mu_a-u_j;\omega_1,\omega_2)
\prod_{b=1}^{s_2} 
\Gamma_h(\nu_b+u_j;\omega_1,\omega_2)}
{\prod_{1\leq i<j\leq n} \Gamma_h(u_i-u_j;\omega_1,\omega_2)
\Gamma_h(u_j-u_i;\omega_1,\omega_2)}
}}
The contour of the integral, which goes from $\IR{\rm e}\, u=-\infty$ to $\IR{\rm e}\, u=+\infty$, 
is chosen appropriately to avoid the poles of the $\Gamma_h$-functions
(see \bult\ for further details).

The function $J_{N_c,(N_f,N_f),0}(\mu;\nu;0)$ that appears in the first degeneration formula
\redab\ expresses the partition function of the $d=3$ $\NN=2$ SQCD theory with gauge group 
$U(N_c)$. When the parameters $\omega_1, \omega_2$ are chosen to have the form
\eqn\redae{
\omega_1 =\i b~, ~~ \omega_2=\i b^{-1}~, ~~ b\in \IR_+
}
this is a partition function on the squashed $S^3$ with squashing parameter $b$ 
\refs{\HamaEA,\ImamuraWG}. The parameters $\mu_a$, $\nu_a$ are related to the real masses 
$m_a$, $\tilde m_a$ and $m_A$ (understood as background values of scalars for 
external vector multiplets of  $SU(N_f)_L$, $SU(N_f)_R$, and $U(1)_A$ respectively) by the 
following relation 
\eqn\redaf{
\mu_a=\tilde m_a+m_A+\omega R_Q~, ~~ \nu_a=-m_a+m_A+\omega R_Q~,~~
\sum_{a=1}^{N_f} m_a=\sum_{a=1}^{N_f} \tilde m_a=0
~.} 
$R_Q$ is the $U(1)_R$ charge of the $d=3$ theory and 
$$
\omega:=\frac{\omega_1+\omega_2}{2}
~.$$
With these conventions the balancing conditions \sciac\ reduce to (\redaf\ sets $x=0$)
\eqn\redag{
\sum_{a=1}^{N_f} \mu_a=\sum_{a=1}^{N_f}\nu_a=
N_f(m_A+\omega R_Q)=\tilde N_c \omega
~.}

Similarly, the second degeneration formula \redac\ expresses the (squashed) $S^3$ partition
function of a $d=3$ $U(N_c)$ SYM theory. If this limit captures correctly the reduction to Aharony
duality, then the partition function on the rhs of eq.\ \redac\ ought to be the partition function of the
magnetic dual of the $d=3$ $\NN=2$ $U(N_c)$ SQCD theory. This is indeed the case. Combining the 
$d=4$ duality transformation property \sciaf\ with the degeneration formulae \redab, \redac\ we obtain 
the identity
\eqn\redai{
J_{N_c,(N_f,N_f),0}(\mu;\nu;0)=
\prod_{a,b=1}^{N_f} \Gamma_h(\mu_a+\nu_b;\omega_1,\omega_2) \
J_{\tilde N_c,(N_f,N_f),0}(\omega-\nu;\omega-\mu;0)
~,}
which is a special case of eq.\ (5.5.21) in Theorem 5.5.11 of Ref.\ \bult\ that expresses Aharony
duality. To the best of our knowledge this particular derivation of the identity \redai\ has not appeared 
in the literature before.

In the next subsection we will see that the balancing condition \redag, which was inherited from four 
dimensions, trivializes the contribution of the gauge-singlet chiral superfields $V_\pm$ that are part of 
the magnetic description of the $d=3$ $\NN=2$ SQCD theory and makes them invisible in the 
degeneration formula \redac. Hence, without relaxing the conditions \redag, it is impossible to read off 
the complete matter content of the magnetic theory. The general form of the identities implied by 
Aharony duality can be obtained by further degeneration limits that remove the conditions \redag.

\subsec{Second degeneration: removal of the balancing conditions and FI terms}

The second degeneration step removes two pairs of quark supermultiplets by sending two pairs
of real masses with opposite signs to infinity. In order to obtain a final theory with $N_f$ 
quark supermultiplets we start from the identity \redai\ renaming
\eqn\redba{
N_f\to N_f+2~, ~~ \tilde N_c\to \tilde N_c+2
~.}
We keep the definition $\tilde N_c=N_f-N_c$ unchanged. In the resulting expression we set 
\eqn\redbba{
\mu_{N_f+1}=\xi_1+\alpha S~, ~~ \nu_{N_f+1}=\zeta_1-\alpha S
~,}
\eqn\redbbb{
\mu_{N_f+2}=\xi_2-\alpha S~, ~~ \nu_{N_f+2}=\zeta_2+\alpha S
}
and eventually take the limit $S\to + \infty$. $\alpha$ is a pure phase chosen in a manner
that allows to perform the ensuing standard reductions by exchanging limits and integrals
\bult. 

With this ansatz the balancing conditions \redag\ become
\eqn\redbc{
\sum_{a=1}^{N_f} \mu_a +\xi_1+\xi_2=
\sum_{a=1}^{N_f} \nu_a+\zeta_1+\zeta_2= 
N_f(m_A+\omega R_Q)=\tilde N_c \omega
}
freeing the parameters $\mu_a, \nu_a$ $(a=1,\ldots,N_f)$ from any constraints.
It will be convenient to define an additional parameter 
\eqn\redbd{
\lambda:=-\xi_1-\zeta_1+\tilde N_c \omega -\frac{1}{2}\sum_{a=1}^{N_f}(\mu_a+\nu_a)
}
in terms of which we obtain the expressions
\eqn\redbe{\eqalign{
\xi_1+\zeta_1=-\lambda+\tilde N_c \omega -\frac{1}{2}\sum_{a=1}^{N_f}(\mu_a+\nu_a)
~,~~
\xi_2+\zeta_2=\lambda+\tilde N_c \omega -\frac{1}{2}\sum_{a=1}^{N_f}(\mu_a+\nu_a)
~.}}
 
Applying the standard reduction identity
\eqn\redbf{\eqalign{
\lim_{S\to \infty} &J_{n,(s_1+2,s_2+2),t}(\mu,\xi_1+\alpha S,\xi_2-\alpha S;
\nu,\zeta_1-\alpha S,\zeta_2+\alpha S;
2 \lambda+\xi_1+\zeta_1-\xi_2-\zeta_2)
\cr
& e^{\frac{\pi \i n}{2\omega_1\omega_2} ((\zeta_1-\alpha S-\omega)^2+
(\xi_2-\alpha S-\omega)^2-(\xi_1+\alpha S-\omega)^2-(\zeta_2+\alpha S-\omega)^2)}
\cr
=&J_{n,(s_1,s_2),t}(\mu;\nu;2 \lambda)
~}}
to the case at hand
\eqn\redbg{
t=0~, ~~ n=N_c~, ~~s_1=s_2=N_f
}
we deduce the limit
\eqn\redbj{\eqalign{
\lim_{S\to \infty} &J_{N_c,(N_f+2,N_f+2),0}(\mu,\xi_1+\alpha S,\xi_2-\alpha S;
\nu,\zeta_1-\alpha S,\zeta_2+\alpha S;0)
\cr
& e^{\frac{\pi \i N_c}{2\omega_1\omega_2} ((\zeta_1-\alpha S-\omega)^2+
(\xi_2-\alpha S-\omega)^2-(\xi_1+\alpha S-\omega)^2-(\zeta_2+\alpha S-\omega)^2)}
\cr
=&J_{N_c,(N_f,N_f),0}(\mu;\nu;2\lambda)
~.}}
The rhs of this equation expresses the (squashed) $S^3$ partition function of the electric
description of the $d=3$ $\NN=2$ $U(N_c)$ SQCD theory with FI term $\lambda$ and 
no restrictions on the real mass parameters $\mu_a, \nu_a$. This explains how the degeneration
limit $S\to \infty$ acts on the lhs of the duality relation \redai. 

The effect of the limit on the magnetic side of the duality follows by inserting the transformation
\redai\ into the lhs of the reduction formula \redbj
\eqn\redbk{\eqalign{
&\lim_{S\to \infty} \Bigg(
\prod_{a,b=1}^{N_f+2} \Gamma_h(\mu_a+\nu_b;\omega_1,\omega_2)
\cr
&J_{\tilde N_c+2,(N_f+2,N_f+2),0}(\omega-\nu,\omega-\zeta_1+\alpha S,\omega-\zeta_2-\alpha S;
\omega-\mu,\omega-\xi_1-\alpha S,\omega-\xi_2+\alpha S;0)
\cr
&e^{\frac{\pi \i N_c}{2\omega_1\omega_2} ((\zeta_1-qS-\omega)^2+
(\xi_2-qS-\omega)^2-(\xi_1+qS-\omega)^2-(\zeta_2+qS-\omega)^2)} \Bigg)
\cr
&=Z_M(\tilde N_c,N_f;\mu;\nu;\lambda) e^{\frac{\pi \i \lambda}{\omega_1\omega_2}
\sum_{a=1}^{N_f}(\mu_a-\nu_a)}
~.}}
We have denoted the result of this limit by using a function $Z_M$. It is clear that $Z_M$ cannot
be obtained with the application of the standard reduction formula \redbf. That formula would 
reduce to the function $J_{\tilde N_c+2, (N_f,N_f),0}$ keeping the number of integration variables
invariant. This is in direct contradiction with the basic duality formula $\tilde N_c=N_f-N_c$, which 
is already apparent from the duality identity \redai. We conclude that the degeneration limit \redbk\
is mathematically more involved than the standard one in \redbf\  and cannot be obtained by 
exchanging limits and integrals. We will call such degeneration limits `non-standard' to distinguish 
them from the standard ones that play a prominent role in the degeneration schemes of Ref.\ \bult. 
Unfortunately, we are not aware of an efficient computational method for such limits, but
we will have more to say about them in the next section.

It is mathematically interesting that the alternate degeneration scheme of Ref.\ \bult\ allows us to 
bypass this complicated reduction formula and derive the function $Z_M$ by using a significantly
different scheme based only on standard reductions.\foot{There is no a priori reason to anticipate
that this alternate route will be a generic possibility. We expect non-standard reductions, like the
one above, to be one of the main steps in general reductions of $d=4$ dualities to $d=3$ dualities.
The example of $SU(N)$ dualities in section 4 appears to be an illustration of this statement.} 
The result, which follows from eq.\ (5.5.21) in Theorem 5.5.11 of \bult, determines $Z_M$ as the dual 
of the rhs of eq.\ \redbj
\eqn\redbl{\eqalign{
Z_M(\tilde N_c, N_f;\mu;\nu;\lambda)=&
\Gamma_h\left( (\tilde N_c+1)\omega-\frac{1}{2}\sum_{a=1}^{N_f}(\mu_a+\nu_a)\pm \lambda\right)
\prod_{a,b=1}^{N_f} \Gamma_h(\mu_a+\nu_b) 
\cr
&J_{\tilde N_c, (N_f,N_f),0}(\omega-\nu;\omega-\mu;-2\lambda)
~.}}
The first $\Gamma_h$ factor on the rhs of this equation captures the contribution of the 
gauge-singlet multiplets $V_\pm$. This factor is invisible in the special case of the balancing 
condition \redag\ since 
\eqn\redam{
\Gamma_h^2\left( (\tilde N_c+1)\omega-\frac{1}{2}\sum_{a=1}^{N_f} (\mu_a+\nu_a)\right)
=\Gamma_h^2 (\omega)=1
~.}
The second term, which is a product of $\Gamma_h$-functions, captures the contribution of the
$N_f^2$ gauge-singlet meson superfields of the dual description. The contribution of the dual
gauge fields and quarks comes into the last factor $J_{\tilde N_c,(N_f,N_f),0}$.

\subsec{Third degeneration: Chern-Simons interactions}

There is a standard third reduction which corresponds to integrating out real masses 
with the same sign. This operation introduces the Chern-Simons interaction. The resulting 
Chern-Simons-matter theories exhibit the Giveon-Kutasov duality \GiveonZN. Since this is a well 
known standard step we will not discuss it explicitly here. For completeness and later convenience 
we list the (squashed) $S^3$ partition functions for the electric and magnetic descriptions of the 
$d=3$ $\NN=2$ $U(N_c)$ Chern-Simons theory at level $k$ coupled to $N_f$ pairs of 
(anti)fundamental supermultiplets, and the duality transformation property that relates them. Without 
loss of generality we assume that the level $k$ is positive.

The electric and magnetic partition functions have respectively the following forms
\eqn\redca{
Z_E(N_c,N_f,k;\mu;\nu;\lambda)=J_{N_c,(N_f,N_f),2k}(\mu;\nu;2\lambda)
~,}
\eqn\redcb{
Z_M(\tilde N_c,N_f,k;\mu;\nu;\lambda)=\prod_{a,b=1}^{N_f} \Gamma_h(\mu_a+\nu_b)\,
J_{\tilde N_c,(N_f,N_f),-2k}(\omega-\nu;\omega-\mu;-2\lambda)
~.}
$\lambda$ denotes again a FI term and $\mu,\nu$ are vectors of real mass parameters.
The Giveon-Kutasov duality requires the transformation property
\eqn\redcc{
Z_E(N_c,N_f,k;\mu;\nu;\lambda)=e^{\i \vartheta(N_c,N_f,k;\mu;\nu;2\lambda)} 
Z_M(\tilde N_c,N_f,k;\mu;\nu;\lambda)
}
where
\eqn\redcd{\eqalign{
&e^{\i \vartheta(N_c,N_f,k;\mu;\nu;\lambda)}:=
e^{\frac{\pi \i(\omega_1^2+\omega_2^2)(k^2+2)}{24\omega_1\omega_2}}
e^{-\frac{\pi \i}{4\omega_1\omega_2}(\lambda^2+2k\omega^2(\tilde N_c-N_c))}
\cr
&e^{\frac{\pi \i}{4\omega_1\omega_2}(2k\sum_a(\omega-\mu_a)^2+2k\sum_a(\omega-\nu_a)^2
-(2(N_c-N_f)\omega+\sum_a\mu_a +\sum_a \nu_a)^2)}
\cr
&e^{-\frac{\pi \i}{2\omega_1\omega_2}(\lambda(\sum_a \nu_a -\sum_a \mu_a)
+2k\omega(2N_f\omega-\sum_a \mu_a-\sum_a \nu_a))}
~.}}
Eq.\ \redcc\ is the last identity of Theorem 5.5.11 in \bult\ as was already noticed in \WillettGP.

\newsec{A lesson from Giveon-Kutasov duality tests}

In the original work on Seiberg duality in $\NN=2$ Chern-Simons-matter theories \GiveonZN\
several checks were performed on the duality using a D-brane setup of D3, D5, NS5 and $(1,k)$
fivebrane bound states. One of these checks aimed to verify that the duality is consistent with the limit
where equal masses with opposite sign for a quark pair $(Q^1, \tilde Q_1)$ are sent to infinity 
removing the respective supermultiplets. Since this limit is a clean example of the non-standard
reduction that we discussed in the previous section, it will be instructive to consider it here in more
detail from the partition function point of view.

An interesting feature of this reduction, that was also noted in \GiveonZN, is the fact that it involves
two separate supersymmetric vacua. In other words, the reduction can be performed in two 
different ways. In vacuum 1, $N_c \to N_c$, $N_f \to N_f-1$ on the electric side; on the magnetic
side $\tilde N_c \to \tilde N_c-1$, $N_f\to N_f-1$. In vacuum 2, $N_c \to N_c-1$, $N_f\to N_f-1$
on the electric side, and $\tilde N_c \to \tilde N_c$, $N_f\to N_f-1$ on the magnetic side.
We notice that there is a possibility of two different types of reductions:
a standard one that keeps the rank of the gauge group (equivalently, the number of
integration variables in the $S^3$ partition function) invariant, and a non-standard one that 
changes the rank of the gauge group. 

In the D-brane interpretation of the duality, \GiveonZN, the following branes in $\IR^{1,9}$
participate
\eqn\GKaaa{\eqalign{
\vbox{ \offinterlineskip  \halign
{ # & #  & #  &  #  & #  &  # &  #  &  #  & #  & # & #   \cr
      &  0 & 1  & 2   & 3  & 4  & 5   & 6   & 7  & 8 & 9   \cr 
    \strut &&&&&&&&&&\cr
1 NS5~:~  & $\bullet$ & $\bullet$ & $\bullet$ & $\bullet$ & $\bullet$ & $\bullet$ &  &&&    \cr
  \strut &&&&&&&&&&\cr
1 $(1,k)$~:~  & $\bullet$  & $\bullet$ & $\bullet$ & $\circ$ &  &  & & $\circ$ & $\bullet$ & $\bullet$    \cr
  \strut &&&&&&&&&&\cr
$N_c$ D3~:~  & $\bullet$ & $\bullet$ & $\bullet$ &  &  & & $\bullet$ &&&    \cr
  \strut &&&&&&&&&&\cr
$N_f$ D5~:~  & $\bullet$ & $\bullet$ & $\bullet$ & &  &  &  & $\bullet$ & $\bullet$ & $\bullet$     \cr
      }
}}}
The circles $\circ$ indicate that the brane is oriented along a line in the $(37)$ plane. 
Giving equal and opposite real masses to a quark pair corresponds to moving the corresponding 
D5 brane away from the D3 branes in the 3-direction. In vacuum 1 the D3 branes continue to 
stretch between the NS5 brane and the $(1,k)$ fivebrane. In vacuum 2 the D3 branes break 
on the displaced D5 brane.

On the level of the $S^3$ partition functions \redca, \redcb\ we set 
\eqn\GKaa{
\mu_1=\xi+\alpha S~, ~~ \nu_1=\zeta-\alpha S
}
and eventually take the limit $S \to +\infty$. This is a slightly simpler version of the reductions 
\redbba, \redbbb\ in subsection 2.3. In the absence of balancing conditions it is now possible to 
consider the limit on a single quark-antiquark supermultiplet pair. 

\subsec{Vacuum 1}

We perform the standard reduction on the electric side 
\eqn\GKab{\eqalign{
&\lim_{S\to \infty} J_{N_c,(N_f,N_f),2k}
(\mu,\xi+\alpha S;\nu,\zeta-\alpha S;2\lambda+\xi+\zeta-2\omega)
e^{\frac{\pi \i N_c}{2\omega_1\omega_2}((\zeta-\alpha S-\omega)^2-(\xi+\alpha S-\omega)^2)}
\cr
&=J_{N_c,(N_f-1,N_f-1),2k}(\mu;\nu;2\lambda)
~.}}
This degeneration formula is the second formula in Proposition 5.3.24 of Ref.\ \bult\ for 
$\tau=\omega$. It holds with certain assumptions on the external parameters $(\mu,\nu)$, 
$(\xi,\zeta)$, and $\varphi={\rm arg}(\alpha)$, which are listed in \bult. The assumption we 
want to single out is the assumption on $\varphi$
\eqn\GKac{
\varphi \in \left( \varphi_+-\pi~, \frac{\varphi_-+\varphi_+ -\pi}{2} \right) \cap
\left( \varphi_\omega -\pi, \varphi_\omega \right)
~,}
where
\eqn\GKad{
\varphi_\omega=\arg (\omega)~, ~~
\varphi_+=\max(\arg (\omega_1),\arg (\omega_2))~, ~~
\varphi_-=\min(\arg(\omega_1),\arg (\omega_2))
~.}
In the case of physical interest \redae\
\eqn\GKae{
\varphi_-=\varphi_+=\varphi_\omega=\frac{\pi}{2} ~~~{\rm and}~~~
\varphi\in \left( -\frac{\pi}{2},0\right)
~.}
The constraint \GKac\ restricts the direction along which we take the limit and is instrumental
when we exchange the limit and integral to derive the degeneration formula \GKab.

The corresponding action on the magnetic side follows from \GKab\ with the use of the
transformation identity \redcc\ on both sides of the equation
\eqn\GKae{\eqalign{
&\lim_{S\to \infty} \Bigg[
e^{\i \vartheta(N_c,N_f,k;\mu,\xi+\alpha S;\nu,\zeta-\alpha S;2\lambda+\xi+\zeta-2\omega)}
e^{\frac{\pi \i N_c}{2\omega_1\omega_2}((\zeta-\alpha S-\omega)^2-(\xi+\alpha S-\omega)^2)}
\cr
&Z_M\left(\tilde N_c,N_f,k;\mu,\xi+\alpha S;\nu,\zeta-\alpha S;
\lambda+\frac{1}{2}(\xi+\zeta)-\omega\right)
\Bigg]
\cr
&= e^{\i \vartheta(N_c,N_f-1,k;\mu;\nu;2\lambda)}
~Z_M(\tilde N_c-1,N_f-1,k;\mu;\nu;\lambda)
}}
giving the expected reduction of the rank of the dual gauge group $\tilde N_c\to \tilde N_c-1$.

Formula \GKae\ is a clean example of what we call a non-standard reduction. Notice, that as soon as 
we assume the validity of \GKac\ in order to implement \GKab\ on the electric side, we no longer have 
the option of a standard reduction, where we exchange limit and integral, on the magnetic side. 
Indeed, that option on the magnetic side \GKae\ would require taking in addition 
\eqn\GKaf{
\varphi \in \left( \frac{\varphi_- +\varphi_+ -\pi}{2},\varphi_- \right) \cap 
\left( \varphi_\omega-\pi, \varphi_\omega \right)
}
which has zero intersection with \GKac. Hence, we are forced uniquely by duality to reduce on the 
magnetic side along the lines of eq.\ \GKae.

\subsec{Vacuum 2}

In this case we make a different choice. We adopt \GKaf\ and perform a standard reduction on 
the magnetic side 
\eqn\GKdd{\eqalign{
\lim_{S\to \infty} \Bigg[ &
e^{\frac{\pi \i \tilde N_c}{2\omega_1\omega_2}((\xi+\alpha S)^2-(\zeta-\alpha S)^2)}
\prod_{a=2}^{N_f} e^{-\frac{\pi \i }{2\omega_1\omega_2}
((\xi+\alpha S+\nu_a)^2-(\zeta-\alpha S+\mu_a)^2)} 
\cr
&Z_M\left( \tilde N_c,N_f,k;\mu,\xi+\alpha S;\nu,\zeta-\alpha S
;\lambda+\frac{1}{2}(\xi+\zeta) \right)\Bigg]
\cr
&=\Gamma_h(\xi+\zeta) Z_M(\tilde N_c,N_f-1,k;\mu,\nu;\lambda)
~. }}
Then, combining this formula with the duality relation we obtain a new non-standard 
reduction on the electric side
\eqn\GKde{\eqalign{
&\lim_{S \to \infty} \Bigg[ 
e^{\frac{\pi \i \tilde N_c}{2\omega_1\omega_2}((\xi+\alpha S)^2-(\zeta-\alpha S)^2)}
\prod_{a=2}^{N_f} e^{-\frac{\pi \i}{2\omega_1\omega_2}((\xi+\alpha S+\nu_a)^2-
(\zeta-\alpha S+\mu_a)^2)}
\cr
&e^{-\i \vartheta(N_c,N_f,k;\mu,\xi+\alpha S;\nu,\zeta-\alpha S;2\lambda+\xi+\zeta)}
Z_E\left( N_c,N_f,k;\mu,\xi+\alpha S;\nu,\zeta-\alpha S;\lambda+\frac{1}{2}(\xi+\zeta) \right) \Bigg]
\cr
&=e^{-\i \vartheta(N_c-1,N_f-1,k;\mu;\nu;2\lambda)}
\Gamma_h(\xi+\zeta) Z_E(N_c-1,N_f-1,k;\mu;\nu;\lambda)
~.}}

The factor $\Gamma_h(\xi+\zeta)$ also comes out in accordance with the D-brane picture.
As we mentioned above, in vacuum 2 the D3 branes break on the displaced D5 brane. Hence, 
as the D5 moves away along $x^3$, half of the broken D3 stretches between the NS5 and the D5, 
while the other half stretches between the D5 and the $(1,k)$ fivebrane. The latter half gives rise to 
a meson supermultiplet degree of freedom that decouples from the rest of the theory. The extra factor 
$\Gamma_h(\xi+\zeta)$ in the rhs of eqs.\ \GKdd-\GKde\ accounts correctly for this additional 
decoupled degree of freedom.

\newsec{Towards 3d Seiberg duality with $SU(N)$ gauge group}

By gauging the baryon symmetry $U(1)_B$ of the $d=4$ $\NN=1$ SQCD theory and then reducing
its SCI we recovered the well known partition function identities required by the Aharony and 
Giveon-Kutasov dualities for the $d=3$ $\NN=2$ SQCD theories with gauge group $U(N_c)$.
Similar dualities for the $d=3$ $\NN=2$ SQCD theory with gauge group $SU(N_c)$ 
(in the presence or absence of Chern-Simons interactions)
have long been suspected to exist, but a viable proposal has not been proposed so far. 
The general philosophy of this work suggests the following approach to this problem. 

Reducing the SCI \sciaa\ of the four dimensional SQCD theory without gauging the baryon symmetry
we obtain the (squashed) $S^3$ partition function of the $d=3$ $\NN=2$ SQCD theory
with gauge group $SU(N_c)$ (see also Theorem 4.6 of \Rainslimits)
\eqn\suaa{
\lim_{v\to 0}I_E^{(SU)}(N_c,N_f;s;t;p,q)
=e^{\frac{\pi \i\omega(N_c^2+1)}{6v\omega_1\omega_2}}
\tilde J_{N_c,(N_f,N_f),0}(\mu;\nu)
}
where we have defined
\eqn\suab{\eqalign{
&\tilde J_{N_c,(N_f,N_f),0}(\mu;\nu)=
\cr
&\frac{1}{N_c!}
\int \prod_{j=1}^{N_c-1}\frac{du_j}{\sqrt{-\omega_1\omega_2}}
\frac{\displaystyle \prod_{a=1}^{N_f} \prod_{j=1}^{N_c} \Gamma_h(\mu_a-u_j;\omega_1,\omega_2)
\Gamma_h(\nu_a+u_j;\omega_1,\omega_2)}
{\displaystyle \prod_{1\leq i<j\leq N_c}\Gamma_h(u_i-u_j;\omega_1,\omega_2)
\Gamma_h(u_j-u_i;\omega_1,\omega_2)}\Bigg |_{\sum_{j=1}^{N_c} u_j=0}
~.}}

A similar reduction on the magnetic description of the four dimensional theory gives
\eqn\suad{
\lim_{v\to 0}I_M^{(SU)}(\tilde N_c,N_f;s;t;p,q)=
e^{\frac{\pi \i\omega(N_c^2+1)}{6v\omega_1\omega_2}}
\prod_{a,b=1}^{N_f} \Gamma_h(\mu_a+\nu_b;\omega_1,\omega_2)
\tilde J_{\tilde N_c,(N_f,N_f),0}(\omega-\mu;\omega-\nu)
~.}
Consequently, the duality identity \sciae\ implies 
\eqn\suac{
\tilde J_{N_c,(N_f,N_f),0}(\mu;\nu)=
\prod_{a,b=1}^{N_f} \Gamma_h(\mu_a+\nu_b;\omega_1,\omega_2)
\tilde J_{\tilde N_c,(N_f,N_f),0}(\omega-\mu;\omega-\nu)
~.}

The balancing conditions \sciac\ reduce to
\eqn\suae{ 
ST^{-1}=(pq)^{\tilde N_c} ~~ \Rightarrow ~~ \sum_{a=1}^{N_f} (\mu_a +\nu_a)=2\tilde N_c \omega
~.}
This condition does not allow a straightforward field theory interpretation of the duality relation \suac. 
A second degeneration step based on the ansatz \redbba, \redbbb\ can be applied on both
sides to remove the balancing condition and lead to the $SU(N_c)$ version of Aharony duality.
On the electric side this is a straightforward standard reduction. On the magnetic side this is a
non-standard reduction. Unfortunately, the absence of an efficient computational method for such 
reductions hinders the completion of this exercise. The result would allow us to determine the 
(squashed) $S^3$ partition function of the magnetic theory from which its full matter content can be 
determined. A third standard reduction that sends equal real masses of the same sign to infinity
can then be used to determine the $SU(N_c)$ version of the Giveon-Kutasov duality.

\newsec{Other features of degeneration schemes}

In this paper we discussed explicitly on the level of SCIs and $S^3$ partition functions
how the standard Seiberg duality for the four dimensional $\NN=1$ SQCD theory reduces in three 
dimensions to Aharony and Giveon-Kutasov dualities for $\NN=2$ SQCD theories. On a 
mathematical level we argued that this reduction entails a set of non-standard degeneration
identities which cannot be determined by exchanging limits and integrals. An efficient 
method for the computation of such identities remains an open problem. On a physical level 
we proposed that such degenerations can be used to determine the precise properties of the
still illusive Aharony and Giveon-Kutasov dualities for $d=3$ $\NN=2$ SQCD theories with 
$SU(N)$ gauge group. Analogous reduction schemes can be envisioned for generic
$d=4$ Seiberg dualities \SpiridonovZA.

In a general 3d/4d connection the degeneration formula
\eqn\concaa{
\lim_{v\to 0} I=Z
}
between a four dimensional SCI $I$ and a three dimensional partition function $Z$ 
can be useful in relating also other properties of the four and three dimensional
theories. An example that deserves further study has to do with spontaneous supersymmetry
breaking. 

If $I$ is zero in a certain regime (independent of fugacities, but dependent on parameters like 
$N_c, N_f$ $etc.$), then according to \concaa\ $Z$ will also be zero in that regime. 
In \MoritaCS\ we conjectured that zeros of $Z$ are related to spontaneous supersymmetry 
breaking in the three dimensional theory. The opposite argument may not be true, namely 
it is not apriori obvious from eq.\ \concaa\ whether $Z=0$ implies $I=0$.

In the specific SQCD example of this note the following spontaneous supersymmetry breaking 
patterns occur. The four dimensional $\NN=1$ SQCD theory exhibits spontaneous supersymmetry 
breaking for $N_f<N_c$. One can readily check (using properties of the elliptic 
$\Gamma_e$-functions) that the SCI index vanishes when $N_f<N_c$ and is non-zero otherwise. 
By reduction the same property carries over to the $S^3$ partition function of the $\NN=2$ SQCD 
theory without CS interactions (for $N_f<N_c$) and the $S^3$ partition function of the $\NN=2$ 
SQCD theory with CS interactions (for $N_f+k<N_c$ and CS level $k$). Without CS interactions it is 
indeed known that a dynamically generated superpotential lifts the space of supersymmetric vacua 
for $N_f<N_c-1$. The case $N_f=N_c-1$ is a bit more tricky, as was already pointed out in \BeniniMF. 
In that case there is a smooth moduli space of supersymmetric vacua. The vanishing of the 
hyperbolic hypergeometric integral expression based on the standard UV description of the theory 
does not reflect this fact, presumably because of accidental symmetries. This subtlety, however, does 
not appear when we make further reductions to obtain Chern-Simons-matter theories. In that case 
spontaneous supersymmetry breaking occurs precisely when the UV description-based $S^3$ 
partition function vanishes.

We believe that such similarities in spontaneous supersymmetry breaking patterns between
four and three dimensional theories related by $S^1$ reductions are naturally explained 
by degeneration formulae of the type \concaa\ in the manner outlined above. Note that the precise 
mechanism of spontaneous breaking of supersymmetry in general depends on the number of 
spacetime dimensions. The details of this connection deserve further study.

\bigskip
\centerline{\bf Acknowledgements}
\medskip

I would like to thank Grigory Vartanov for useful correspondence and discussions. The work of VN 
was partially supported by the European grants FP7-REGPOT-2008-1: CreteHEPCosmo-228644, 
PERG07-GA-2010-268246, and the EU program ``Thalis'' ESF/NSRF 2007-2013.

\listrefs
\end